\documentclass[prl,twocolumn,longbibliography,nofootinbib, preprintnumbers,notitlepage,fleqn,showpacs,superscriptaddress]{revtex4-1}

\usepackage{color,soul}

\usepackage{contour}
\usepackage{ulem}

\contourlength{0.8pt}

\renewcommand{\emph}{\textit}

\usepackage[bookmarks = true, citecolor = blue, colorlinks = true, linkcolor = magenta, urlcolor = blue]{hyperref}

\usepackage[T1]{fontenc}			
\usepackage[sc,osf]{mathpazo}   	

\usepackage{amsmath}  			
\usepackage{amsfonts}  			
\usepackage{graphicx}   			
\usepackage{mathrsfs, amsthm, amssymb}
\usepackage{bbm, bm}

\usepackage{nicefrac}    			

\usepackage{fourier-orns}  		

\usepackage[braket, qm]{qcircuit}	

\usepackage{nicefrac} 

\usepackage{tikz}
\usetikzlibrary{calc,decorations.pathmorphing,shapes}

\newcounter{sarrow}
\newcommand\xrsquigarrow[1]{%
\stepcounter{sarrow}%
\mathrel{\begin{tikzpicture}[baseline= {( $ (current bounding box.south) + (0,-0.5ex) $ )}]
\node[inner sep=.5ex] (\thesarrow) {$\scriptstyle #1$};
\path[draw,<-,decorate,
  decoration={zigzag,amplitude=0.7pt,segment length=1.2mm,pre=lineto,pre length=4pt}] 
    (\thesarrow.south east) -- (\thesarrow.south west);
\end{tikzpicture}}%
}

\def\id{{\rm 1\kern-.22em l}}

\begin{document}


\title[]{Arbitrary entanglement of three qubits via linear optics
}

\author{Pawel \surname{Blasiak}}
\email{pawel.blasiak@ifj.edu.pl}
\affiliation{Institute of Nuclear Physics Polish Academy of Sciences, PL-31342 Krak\'ow, Poland}
\author{Ewa \surname{Borsuk}}
\email{ewa.borsuk@ifj.edu.pl}
\affiliation{Institute of Nuclear Physics Polish Academy of Sciences, PL-31342 Krak\'ow, Poland}

\author{Marcin \surname{Markiewicz}}
\email{marcinm495@gmail.com}
\affiliation{International Centre for Theory of Quantum Technologies, University of Gda\'nsk, PL-80308 Gda\'nsk, Poland}


\begin{abstract}
We present a linear-optical scheme for generation of an arbitrary state of three qubits. It requires only three independent particles in the input and post-selection of the coincidence-type at the output. The success probability of the protocol is equal for any desired state. Furthermore, the optical design remains insensitive to particle statistics (bosons, fermions or anyons). This approach builds upon the no-touching paradigm, which demonstrates the utility of particle indistinguishability as a resource of entanglement for practical applications. 
\end{abstract}


\maketitle

\textit{Introduction.}---Entanglement remains a central theme in the quantum foundations research~\cite{Be93,BrCaPiScWe14}. It is considered as a key resource enabling the advantage in quantum information tasks~\cite{NiCh00,HoHoHoHo09}. There is therefore a vital interest in practical entanglement generation schemes capable of delivering broadest possible range of states, from which one might choose a given state suitable for a problem at hand. A paradigmatic example is the construction of the full class of single particle states (qudit). In this case arbitrary unitary transformation can be experimentally implemented by linear-optical devices~\cite{ReZeBeBe94} and hence any single-particle state can be deterministically prepared from any given initial state. However, it is not true for multi-particle states which do not transform one into another by means of linear optics~\cite{MiRoOsLe14}. Thus for generation of multi-partite entanglement non-linear effects or post-selection of some sort must be employed. Several techniques have been developed to this effect~\cite{PaChLuWeZeZu12,KrMaFiLaZe16,ErKrZe20,WaScLaTh20}, however there is no systematic method for obtaining arbitrary multi-partite entanglement from separable states.

In this work we focus on entanglement generation for three qubits. An interesting approach to this problem consists in considering the set of stochastic local operations and classical communication (SLOCC) \cite{KeLiMa99, DuViCi00, VeDeDe01}. It has been shown that for three qubits there are two inequivalent classes of genuinely tripartite entangled states~\cite{DuViCi00}. This means that using SLOCCs for filtering arbitrary tripartite entanglement requires two different types of entangled states to begin with, i.e., states of the GHZ and W type. It should be remarked that the efficiency of such protocols drops to zero when moving away from the initial state. Another important result concerns the full set of linear optical transformations (without post-selection)~\cite{MiRoOsLe14}. Then the situation is further complicated as for three qubits the set of entangled states splits into a continuous number of inequivalent classes. Those results illustrate the difficulties in generating arbitrary entanglement using linear operations without having any entanglement from the outset.

Here we will show that linear optical transformations augmented with post-selection of the coincidence type is enough to generate arbitrary entanglement of three qubits from three independent particles (i.e., without requiring any prior entanglement). Our proposal builds on the so called \textit{no-touching} paradigm in optical designs which draws from the inherent indistinguishability of particles; see~\cite{BlMa19} for a general scheme and~\cite{YuSt92a,YuSt92,NeOfChHeMaUm07,BoHo13,BlBoMaKi21} for some examples. Notably, the protocol is an instance of a direct and explicit construction of any given state. A distinctive advantage of the proposal is that it has equal efficiency for generation of any desired state and it is insensitive to particle statistics.

\vspace{0.1cm}


\textit{Simplification by generalised Schmidt decomposition.}---A general state of three qubits reads
\begin{eqnarray}\label{psi}
\ket{\psi}&=&\sum_{ijk=0,1}\alpha_{ijk}\ket{ijk},
\end{eqnarray}
where $\ket{ijk}$ is a computational basis in $\mathbb{C}^2\otimes\mathbb{C}^2\otimes\mathbb{C}^2$. It turns out that this can be simplified by local transformations to the combination of the following five states
\begin{eqnarray}\nonumber
\ket{\psi}&=&a\ket{000}+b\,e^{i\varphi}\ket{100}+c\ket{110}+d\ket{101}+e\ket{111},\\\label{psi-reduced}
\end{eqnarray}
with five real parameters $a,b,c,d,e\geqslant0$ and one phase $0\leqslant\varphi\leqslant\pi$, such that
\begin{eqnarray}\label{constraint-0}
a^2+b^2+c^2+d^2+e^2&=&1\,.
\end{eqnarray}
This follows from the generalised Schmidt decomposition for three qubits~\cite{AcAnCoJaLaTa00,CaHiSu00} and no further reduction of the number of non-vanishing terms is possible. 

Therefore, in order to generate the state in Eq.~(\ref{psi}), it is enough to provide state in Eq.~(\ref{psi-reduced}) and then make local transformation on each qubit. In the following we give the explicit protocol for optical construction of an arbitrary state in the reduced form of Eq.~(\ref{psi-reduced}).

\vspace{0.1cm}

\begin{figure*}
\centering
\includegraphics[width=1.2\columnwidth]{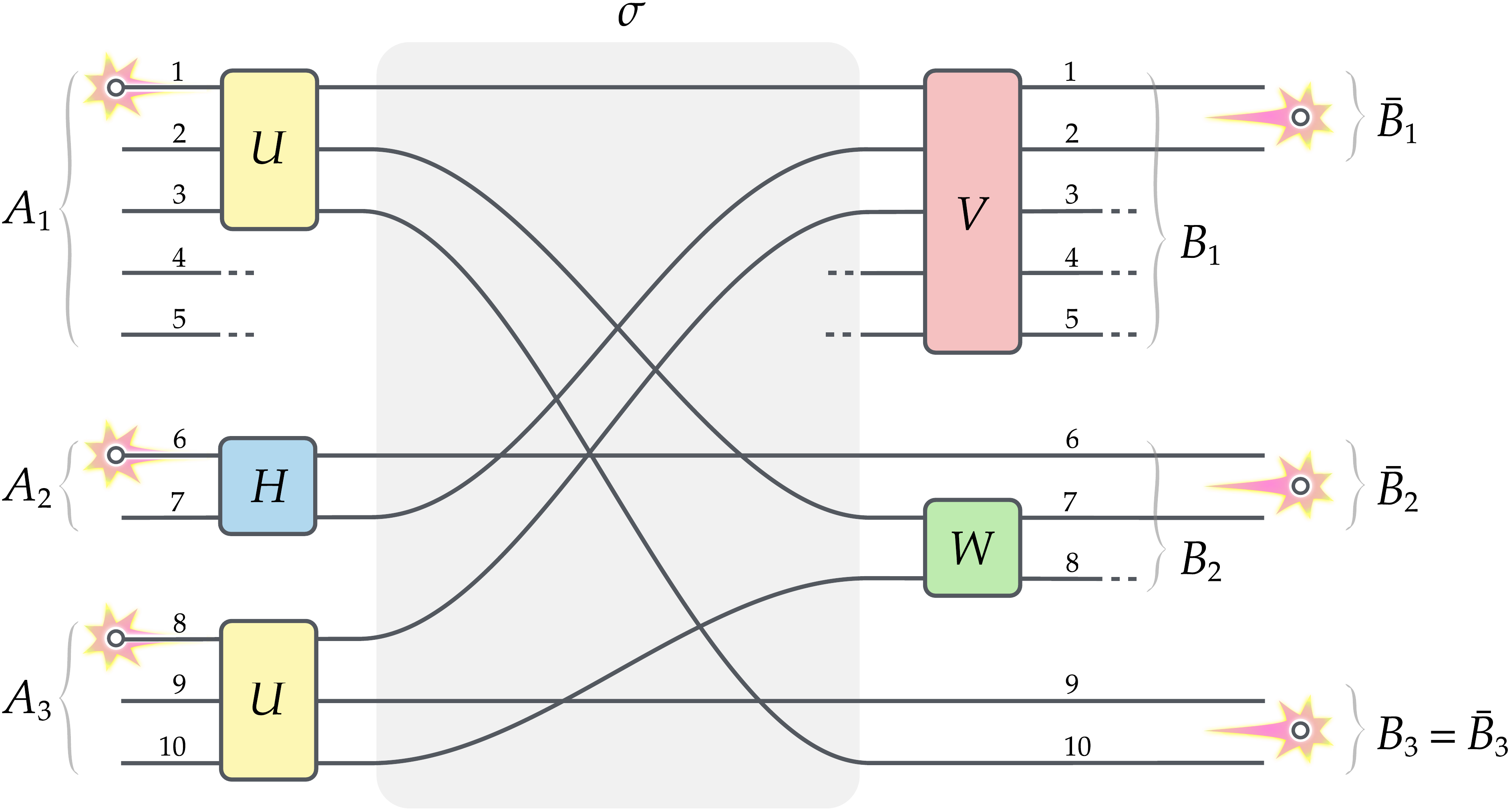}
\caption{\label{Fig_Protocol}{\bf\textsf{\mbox{No-touching design for arbitrary state of three qubits.}}} Three independent identical particles entering the optical circuit undergo a sequence of transformations which consists of  local unitaries $U$, $H$ and $U$ on subsystems $A_1$, $A_2$ and $A_3$, followed by permutation of the paths $\sigma$, and again local unitaries $V$ and $W$ on subsystems $B_1$ and $B_2$. Post-selection on a single particle in each of the dual-rail qubits generates an arbitrary state of three qubits $\bar{B}_1$, $\bar{B}_2$ and $\bar{B}_3$ in Eq.~(\ref{psi-reduced}) by the appropriate choice of unitaries in Eq.~(\ref{VW}) as specified in Eqs.~(\ref{eq-1})-(\ref{eq-3}).}
\end{figure*}

\textit{Optical realisation.}---Consider three identical particles injected into an optical circuit which consists of 10 paths (or modes). The particle statistics (bosons, fermions or anyons) is irrelevant for the purpose at hand. Following the idea presented in Ref.~\cite{BlMa19} we will group paths at the input and output into three systems, see Fig.~\ref{Fig_Protocol}:
\begin{eqnarray}\nonumber
\!\!\!\!\!\!&&\textsl{Input:}\ \ A_1=\{1,2,3,4,5\},\ A_2=\{6,7\},\ A_3=\{8,9,10\},\\\nonumber
\!\!\!\!\!\!&&\textit{Output:}\ \ B_1=\{1,2,3,4,5\},\ B_2=\{6,7,8\},\ B_3=\{9,10\}.
\end{eqnarray}

Furthermore, we will distinguish three pairs of paths $\bar{B}_1$, $\bar{B}_2$ and $\bar{B}_3$ in the respective subsystems at the output, i.e. $\bar{B}_k\subset B_k$. They will play the role of the so called \textit{dual-rail qubits}, where the computational basis $\ket{0}$,$\ket{1}$ is encoded by a single particle in the respective path of a given pair $\bar{B}_k$. Accordingly, we have the following representation for the qubit states $\alpha\ket{0}+\beta\ket{1}$:
\begin{eqnarray}\nonumber
\begin{array}{lcl}
\bar{B}_1=\{1,2\}&\ \ \leadsto\ \ &\textit{Qubit 1:\ \ }(\alpha\, a^\dag_1+\beta\, a^\dag_2)\ket{\Omega},\vspace{0.1cm}\\
\bar{B}_2=\{6,7\}&\ \ \leadsto\ \ &\textit{Qubit 2:\ \ }(\alpha\, a^\dag_6+\beta\, a^\dag_7)\ket{\Omega},\vspace{0.1cm}\\
\bar{B}_3=\{9,10\}&\ \ \leadsto\ \ &\textit{Qubit 3:\ \ }(\alpha\, a^\dag_9+\beta\, a^\dag_{10})\ket{\Omega}.
\end{array}
\end{eqnarray}
where $\ket{\Omega}$ denotes the vacuum state and $a^\dag_i$ are creation operators of the particles, for which we do not specify their commutation relations. We note that the dual-rail encoding assumes the presence of a \textit{single} particle in a given pair of paths $\bar{B}_k$ for $k=1,2,3$. In our scheme this will be guaranteed by \textit{post-selection} on a single particle in each dual-rail qubit $\bar{B}_k$. 

The optical protocol is depicted in Fig.~\ref{Fig_Protocol}. It consists of a sequence of three unitary transformations on three independent particles injected in paths 1, 6 and 8. First, the particles undergo local unitaries $U$, $H$ and $U$ in each subsystem $A_1$, $A_2$ and $A_3$. Second, the paths are rearranged according to some permutation $\sigma\in\mathcal{S}_{10}$. Third, there are two local unitaries $V$ and $W$ on subsystems $B_1$ and $B_2$ implemented at the output. Finally, the protocol ends with post-selection on a single particle in each pair of paths $\bar{B}_1$, $\bar{B}_2$ and $\bar{B}_3$ which generates three dual-rail encoded qubits.


Now, we can make the unitaries in Fig.~\ref{Fig_Protocol} more precise. Let the first two transformations $U$ and $V$ produce symmetric superposition of the injected particles, which in the matrix notation amounts to
\begin{eqnarray}\label{UH}
U\,=\,
\tfrac{1}{\sqrt{3}}\left(\begin{array}{ccc}
1&\textbf{.}&\textbf{.}\\
1&\textbf{.}&\textbf{.}\\
1&\textbf{.}&\textbf{.}\\
\end{array}
\right)\quad\text{\&}\quad H\, =\, 
\tfrac{1}{\sqrt{2}}\left(\begin{array}{rr}
1&\textbf{.}\\
1&\textbf{.}
\end{array}
\right).
\end{eqnarray}
The permutation of modes $\sigma\in\mathcal{S}_{10}$ is given as follows
\begin{eqnarray}\label{sigma}
\sigma&=&\left(\begin{array}{cccccccccc}
1&2&3&4&5&6&7&8&9&10\\
1&7&10&4&5&6&2&3&9&8
\end{array}
\right).
\end{eqnarray}
The final unitaries $V$ and $W$ are defined in a non-trivial way by the following two matrices
\begin{eqnarray}\label{VW}
V\,=\,
\left(\begin{array}{ccccc}
\kappa&0&0&\delta&\epsilon\\
\bar{\delta}&\mu&\nu&\text{-\,}\bar{\kappa}&0\\
\textbf{.}&\textbf{.}&\textbf{.}&\textbf{.}&\textbf{.}\\
\textbf{.}&\textbf{.}&\textbf{.}&\textbf{.}&\textbf{.}\\
\textbf{.}&\textbf{.}&\textbf{.}&\textbf{.}&\textbf{.}
\end{array}
\right)\quad\text{\&}\quad W\,=\,
\left(\begin{array}{cc}
\xi&\tau\\
\textbf{.}&\textbf{.}
\end{array}
\right),
\end{eqnarray}
with some parameters $\kappa,\delta,\nu,\mu,\epsilon,\xi$ and $\tau$.

In the above notation the dots "$\textbf{.}$" are left unspecified and chosen so that the matrices are unitary. Observe that this can be always done by augmenting the missing columns/rows to an orthonormal basis (note that the two upper rows of $V$ are orthogonal at the outset). The only constraint on the parameters $\kappa,\delta,\nu,\mu,\epsilon,\xi$ and $\tau$ is their respective normalisation, i.e.
\begin{eqnarray}\label{constraint-1}
|\xi|^2+|\tau|^2&=&1\,,\\\label{constraint-2}
|\kappa|^2+|\delta|^2+|\epsilon|^2&=&1\,,\\\label{constraint-3}
|\delta|^2+|\mu|^2+|\nu|^2+|\kappa|^2&=&1\,.
\end{eqnarray}
For our purposes the dotted entries "$\textbf{.}$" will play no role in the argument (in the following they contribute only to the terms that drop out upon post-selection). All the remaining parameters  $\kappa,\delta,\nu,\mu,\epsilon,\xi$ and $\tau$ will be specified shortly.

Let us write out the state that results from the protocol in Fig.~\Ref{Fig_Protocol} after injecting three independent particles in paths $1,6,8$ and post-selecting on a single particle in each dual-rail qubit $\bar{B}_1$, $\bar{B}_2$ and $\bar{B}_3$. Evolution of the system is given by a sequence of steps as described in the following lines:

\begin{widetext}
\begin{eqnarray}
\label{evolution-1}
a^{\dag}_{\scriptscriptstyle 1}\,a^{\dag}_{\scriptscriptstyle 6}\,a^{\dag}_{\scriptscriptstyle 8} \, \ket{0}
&\xymatrix{\ar[r]^{\atop U,H,U}_{Eq.\,\,(\ref{UH})\atop }  &}&\tfrac{1}{3\sqrt{2}}\,\big(\,a^{\dag}_{\scriptscriptstyle 1}+a^{\dag}_{\scriptscriptstyle 2}+a^{\dag}_{\scriptscriptstyle 3}\,\big)\,\big(\,a^{\dag}_{\scriptscriptstyle 6}+a^{\dag}_{\scriptscriptstyle 7}\,\big)\,\big(\,a^{\dag}_{\scriptscriptstyle 8}+a^{\dag}_{\scriptscriptstyle 9}+a^{\dag}_{\scriptscriptstyle 10}\,\big)\ket{0}\\
\label{evolution-2}
&\xymatrix{\ar[r]^{\atop \sigma}_{Eq.\,(\ref{sigma})\atop }  &}&\tfrac{1}{3\sqrt{2}}\,\big(\,a^{\dag}_{\scriptscriptstyle 1}+a^{\dag}_{\scriptscriptstyle 7}+a^{\dag}_{\scriptscriptstyle 10}\,\big)\,\big(\,a^{\dag}_{\scriptscriptstyle 6}+a^{\dag}_{\scriptscriptstyle 2}\,\big)\,\big(\,a^{\dag}_{\scriptscriptstyle 3}+a^{\dag}_{\scriptscriptstyle 9}+a^{\dag}_{\scriptscriptstyle 8}\,\big)\ket{0}\\\label{evolution-3}
&\xymatrix{\ar[r]^{\atop V,W}_{Eq.\,(\ref{VW})\atop }  &}&\tfrac{1}{3\sqrt{2}}\,\Big(\big(\,\kappa\,a^{\dag}_{\scriptscriptstyle 1}+\bar{\delta}\,a^{\dag}_{\scriptscriptstyle 2}+...\,\big)+\big(\,\xi\,a^{\dag}_{\scriptscriptstyle 7}+...\,\big)+a^{\dag}_{\scriptscriptstyle 10}\,\Big)\,\\
\label{evolution-3}
&&\qquad\!\Big(\,a^{\dag}_{\scriptscriptstyle 6}+\big(\,\mu\,a^{\dag}_{\scriptscriptstyle 2}+...\,\big)\Big)\\
\label{evolution-4}
&&\qquad\!\Big(\big(\,\nu\,a^{\dag}_{\scriptscriptstyle 2}+...\,\big)+a^{\dag}_{\scriptscriptstyle 9}+\big(\,\tau\,a^{\dag}_{\scriptscriptstyle 7}+...\,\big)\Big)\ket{0}\\
\label{evolution-5}
&\xrsquigarrow{\text{\!\tiny{\emph{post-select}}\!}}&\tfrac{1}{3\sqrt{2}}\,\Big(\,\kappa\,a^{\dag}_{\scriptscriptstyle 1}\,a^{\dag}_{\scriptscriptstyle 6}\,a^{\dag}_{\scriptscriptstyle 9}+\bar{\delta}\,a^{\dag}_{\scriptscriptstyle 2}\,a^{\dag}_{\scriptscriptstyle 6}\,a^{\dag}_{\scriptscriptstyle 9}+\xi\mu\,a^{\dag}_{\scriptscriptstyle 7}\,a^{\dag}_{\scriptscriptstyle 2}\,a^{\dag}_{\scriptscriptstyle 9}+\nu\,a^{\dag}_{\scriptscriptstyle 10}\,a^{\dag}_{\scriptscriptstyle 6}\,a^{\dag}_{\scriptscriptstyle 2}+\mu\tau\,a^{\dag}_{\scriptscriptstyle 10}\,a^{\dag}_{\scriptscriptstyle 2}\,a^{\dag}_{\scriptscriptstyle 7}\,\Big)\ket{0}\\
\label{evolution-6}
&&=\ \tfrac{1}{3\sqrt{2}}\,\Big(\,\kappa\ket{000}+\bar{\delta}\ket{100}+\xi\mu\ket{110}+\nu\ket{101}+\mu\tau\ket{111}\Big)\,,
\end{eqnarray}
\end{widetext}
where the last equality holds for bosons in the dual-rail encoding. 
This renders the desired state in Eq.~(\ref{psi-reduced}) when
\begin{eqnarray}
\kappa=a\,,\ \ \bar{\delta}=b\,e^{i\varphi}\,,\ \ \xi\mu=c\,,\ \ \nu=d\,,\ \ \mu\tau=e\,.
\end{eqnarray}
We note that we can always choose matrices $W$ and $V$ so that these equations are satisfied. It obtains by defining parameters $\kappa,\delta,\nu,\mu,\epsilon,\xi$ and $\tau$ in the following way
\begin{eqnarray}\label{eq-1}
&&\kappa=a\,,\qquad\delta=b\,e^{-i\varphi}\,,\qquad\nu=d\,,\\\label{eq-2}
&&
\mu=\sqrt{c^2+e^2}\,,\qquad\epsilon=\sqrt{1-a^2-b^2}\,,\\\label{eq-3}
&&\xi=\nicefrac{\textstyle{c}}{\textstyle{\mu}}\,,\qquad\tau=\nicefrac{\textstyle{e}}{\textstyle{\mu}}\,.\ \ 
\end{eqnarray}
(If $\mu=0$, then $\xi$ and $\tau$ can be taken arbitrarily). It is straightforward to check that the conditions in Eqs.~(\ref{constraint-1})-(\ref{constraint-3}) are satisfied, since the constraint in Eq.~(\ref{constraint-0}) holds.

Note that the scheme works for fermions as well. In this case the final sate in Eq.~(\ref{evolution-6}) takes the form 
\begin{eqnarray}
\!\!\!\!\!\!\!\!\!\kappa\ket{000}+\bar{\delta}\ket{100}-\xi\mu\ket{110}-\nu\ket{101}+\mu\tau\ket{111},
\end{eqnarray}
and then the Eqs.~(\ref{eq-1})-(\ref{eq-3}) require a trivial modification $\nu=-\,d$ and $\xi=-\,\nicefrac{\textstyle{c}}{\textstyle{\mu}}$ in order to recover the desired state in Eq.~(\ref{psi-reduced}). In a similar manner it is straightforward to adjust Eq.~(\ref{evolution-6}) for \textit{any} particle statistics (anyons). 


Finally, we observe that the expression in Eq.~(\ref{evolution-6}) is unnormalised as a result of post-selection. This allows to read off the success probability (efficiency) of the process which is  equal to $\big(\,\nicefrac{1}{3\sqrt{2}}\,\big)^{\scriptscriptstyle2}=\nicefrac{1}{18}\approx5\,\%$. Notably, the efficiency is the \underline{\textit{same}} for \textit{every} three qubit state $\ket{\psi}$.

\vspace{0.1cm}

\begin{figure}
	\centering
\includegraphics[width= \columnwidth ]{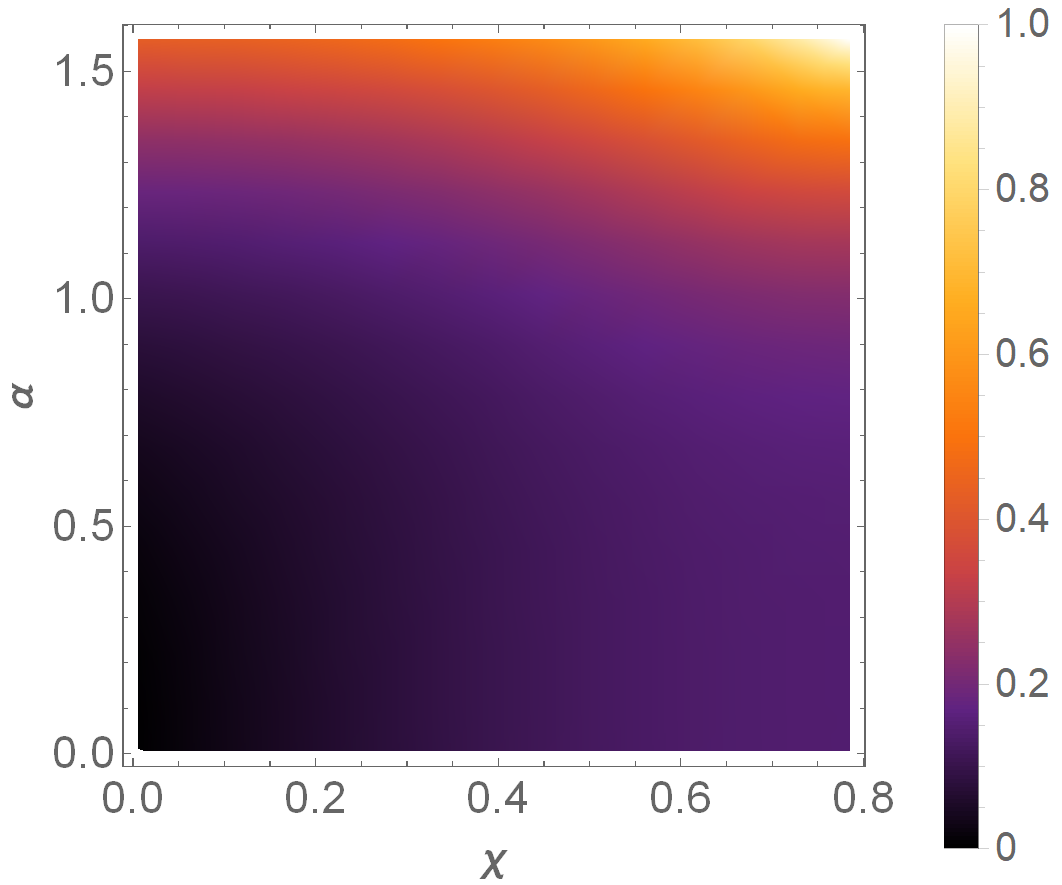}
	\caption{\label{FigGHZ}
  Success probability of obtaining arbitrary state from a two-parameter subclass of the GHZ class in Eq.~\eqref{GHZclass} of the form $\ket{\psi_{\textrm{GHZ}}(\chi,\pi,\alpha,\alpha,\alpha)}$. For $\chi=\frac{\pi}{4}$ and $\alpha\rightarrow 0$ the success probability vanishes, which indicates that these states cannot be effectively obtained via SLOCC filtering.
   } 
\end{figure}

\textit{Comparison with generation of arbitrary states via SLOCC operations.} In the previous section we have presented a universal interferometric protocol for generation of arbitrary three-qubit state from an input product state of three particles, which generates arbitrary state with constant finite efficiency. In this section we would like to compare this scheme with generation of arbitrary state from a GHZ class starting from the GHZ input state via SLOCC operations. We will see that this method of state generation has a vanishing efficiency for some states in this class. As shown in the seminal paper by Dur et. al.~\cite{DuViCi00} arbitrary state from the GHZ class can be parametrised by five real parameters as
\begin{eqnarray}
\label{GHZclass}
&&\ket{\psi_{\textrm{GHZ}}(\chi,\theta,\alpha_1,\alpha_2,\alpha_3)}=\nonumber\\
&&\sqrt{K}\left(\cos(\chi)\ket{000}+\sin(\chi)e^{i\theta}\ket{s_1}\ket{s_2}\ket{s_3}\right),\nonumber\\
\end{eqnarray}
in which the normalisation constant reads $K=(1+2\cos(\chi)\sin(\chi)\cos(\alpha_1)\cos(\alpha_2)\cos(\alpha_3)\cos(\theta))^{-1}$ and the states $\ket{s_i}$ are given by $\cos(\alpha_i)\ket{0}+\sin(\alpha_i)\ket{1}$. The ranges of the parameters are as follows: $\chi\in (0,\frac{\pi}{4}]$, $\alpha_i\in(0, \frac{\pi}{2}]$ and $\theta\in[0,2\pi)$.
This state can be obtained from the standard GHZ state $\ket{\textrm{GHZ}}=\frac{1}{\sqrt 2}(\ket{000}+\ket{111})$ via SLOCC filtering operations specified by
\begin{equation}
\ket{\psi_{\textrm{GHZ}}(\chi,\theta,\alpha_1,\alpha_2,\alpha_3)}=M(\chi,\theta,\alpha_1,\alpha_2,\alpha_3)\ket{\textrm{GHZ}},
\end{equation}
where the SLOCC operator $M$ has the form~\cite{DuViCi00}
\begin{eqnarray}
\label{MGen}
M(\chi,\theta,\alpha_1,\alpha_2,\alpha_3)&=&\sqrt{2K}\left(\begin{array}{rr}
\cos(\chi)&\sin(\chi)\cos(\alpha_1)e^{i\theta}\\
0&\sin(\chi)\sin(\alpha_1)e^{i\theta}
\end{array}
\right)\nonumber\\
&&\otimes\left(\begin{array}{rr}
1&\cos(\alpha_2)\\
0&\sin(\alpha_2)
\end{array}
\right)
\otimes\left(\begin{array}{rr}
1&\cos(\alpha_3)\\
0&\sin(\alpha_3)
\end{array}
\right)\nonumber\\
&=&\sqrt{2K}\,\tilde{M}(\chi,\theta,\alpha_1,\alpha_2,\alpha_3).
\end{eqnarray}
Such a filtering operation can be implemented as a two-outcome POVM measurement~\cite{VeDeDe01}, with measurement operators defined by $P=M/||M||$ and $P'=\sqrt{\id-P^{\dagger}P}$. The outcome related with the measurement operator $P$ indicates success of the protocol, whereas the outcome related with $P'$ --- its failure. The norm has to be chosen in a way which guarantees that $P^{\dagger}P\leq\id$. One of the typical choices is the spectral norm of the operator $M$ defined as the largest singular value of $M$. This choice turns out to be optimal for the task of entanglement distillation of two-qubit states~\cite{VeDeDe01}, however other choices, which guarantee the condition 
$P^{\dagger}P\leq\id$, as for example the Frobenius norm, are also correct.
The success probability of filtering arbitrary state of the form Eq.~\eqref{GHZclass} for SLOCC operator $M$ is thus given by~\cite{AvKe09}
\begin{eqnarray}
\label{pSucc}
p_{\textrm{succ}}&=&\operatorname{Tr}(P\rho_{\textrm{GHZ}}P^{\dagger})=\operatorname{Tr}\left(\frac{M}{||M||}\rho_{\textrm{GHZ}}\frac{M^{\dagger}}{||M^{\dagger}||}\right)\nonumber\\
&=&\frac{\operatorname{Tr}(M\rho_{\textrm{GHZ}}M^{\dagger})}{||M||^2}=\frac{1}{||M||^2},
\end{eqnarray}
in which $\rho_{\textrm{GHZ}}=\ket{\textrm{GHZ}}\bra{\textrm{GHZ}}$, and the last equality follows from the fact that $\ket{\psi_{\textrm{GHZ}}}$ is already properly normalised. Note that the operator $M$ is \textit{not} unitary, and therefore it does not preserve normalisation of a general state it acts on --- the state $\rho_{\textrm{GHZ}}$ is an exception. Let us firstly assume that we choose the spectral norm in Eq.~\eqref{pSucc}.
For the clarity of presentation let us focus on a two-parameter subclass of states from the GHZ class Eq.~\eqref{GHZclass} of the form $\ket{\psi_{\textrm{GHZ}}(\chi,\pi,\alpha,\alpha,\alpha)}$. In the Fig.~\ref{FigGHZ} we present the success probability of obtaining this state from a GHZ state as a function of parameters $\chi$ and $\alpha$. We can see that the probability tends to zero for $\chi=\frac{\pi}{4}$ and $\alpha\rightarrow 0$, which stands in sharp contrast with our protocol, that allows for generation of these states with fixed finite probability of success independently of the values of the parameters. One may argue that the effect of vanishing probability is related with a specific choice of the norm. However, it is easy to see that this effect holds for any choice of the norm consistent with the condition $P^{\dagger}P\leq\id$. Indeed, it suffices to show that $||M(\tfrac{\pi}{4},\pi,\alpha\rightarrow 0)||\rightarrow \infty$ for any choice of the norm. Due to Eq.~\eqref{MGen} we have
\begin{eqnarray}
   &&||M(\tfrac{\pi}{4},\pi,\alpha\rightarrow0)||=\nonumber\\
   &&\left|\sqrt{2K\left(\tfrac{\pi}{4},\pi,\alpha\rightarrow 0\right)}\right|\cdot ||\tilde{M}(\tfrac{\pi}{4},\pi,\alpha\rightarrow 0)||. 
\end{eqnarray}
Now it can be easily directly verified that $\left|\sqrt{2K\left(\tfrac{\pi}{4},\pi,\alpha\rightarrow 0\right)}\right|\rightarrow\infty$. Therefore it suffices to show that $||\tilde{M}(\tfrac{\pi}{4},\pi,\alpha\rightarrow 0)||$ is strictly positive for any choice of the norm. For spectral norm one has $||\tilde{M}(\tfrac{\pi}{4},\pi,\alpha\rightarrow 0)||=2$. All  matrix norms for finite dimensional matrices of a fixed dimension are equivalent, which means that for any two norms $||\cdot||_X$ and $||\cdot||_Y$ there exist two \textit{positive} numbers
$x,x'$ such that for any matrix $A$ one has $x||A||_X\leq ||A||_Y\leq x'||A||_X$.
From this property it follows that $||\tilde{M}(\tfrac{\pi}{4},\pi,\alpha\rightarrow 0)||$ must be strictly positive for any choice of the matrix norm, which implies $||M(\tfrac{\pi}{4},\pi,\alpha\rightarrow 0)||\rightarrow \infty$ and therefore the success probability for filtering the states in the neighbourhood of $\chi=\tfrac{\pi}{4}$ and $\alpha=0$ for any implementation of the SLOCC operation \eqref{MGen} is arbitrarily close to zero.

This shows that our protocol overcomes the difficulties of a state generation via SLOCC filtering operations, as there: (i) the filtering probability can be vanishing, (ii) we are confined within one of the six entanglement classes depending on the initial state of the filtering. Both restrictions do not occur in our protocol, as the success probability is constant for any state and one can reach arbitrary state regardless of its entanglement class from the same trivial initial state.

\vspace{0.1cm}


\textit{Discussion.}---We remark that the above described protocol, based on dual-rail encoded qubits, serves as a template that can be straightforwardly translated into any other physical implementation of qubits. This is a generic feature of the no-touching designs in which the question of particle statistics becomes virtually irrelevant due to post-selection~\cite{BlMa19,YuSt92a,YuSt92,NeOfChHeMaUm07,BoHo13,BlBoMaKi21}. 

From the fundamental point of view, it is interesting to note the significance of the inherent indistinguishability of particles as conveniently described in the second quantisation formalism. It appears that entanglement resulting from the symmetrization postulate can be treated as a genuine resource and transformed into other kinds of entanglement which can be directly observed and used for practical applications~\cite{BlMa19,LoCo16,LoCo18,BlBoMa21}. This paper shows that arbitrary entanglement of three qubits can be extracted in this way.

An important advantage of the proposed protocol is the minimal amount of resources employed to generate an arbitrary three-qubit state compared to the existing techniques. It requires only linear optics and works equally well for any particle statistics, cf.~\cite{ PaChLuWeZeZu12,KrMaFiLaZe16,ErKrZe20,WaScLaTh20}. There is no need for auxiliary systems, particles, or measurements, cf.~\cite{BeLoCo17,KiChLiHa20}. Furthermore, the protocol requires only three independent particles in the input, i.e. no prior entanglement is required, and it has the same efficiency for generation of any desired state. This distinguishes our proposal from the typical approach based on filtering via SLOCC operations which requires auxiliary entanglement from the outset and its success probability for arbitrary three-qubit states drops to zero.Moreover generation of states via SLOCC filtering demands in general different initial states depending on the SLOCC equivalence class of the target state. For optical proposals aimed at preparation of single representatives in the SLOCC classes for the purpose of filtering see Refs.~\cite{BlMa19,KiPrChYaHaLeKa18,JuYaPaChCa19,LePrChLiChKi21}. Notably, our protocol overcomes the division into SLOCC equivalence classes due to the presence of mode permutation $\sigma$, which is non-local operation from the point of view of subsystems defined by mode grouping.

In summary, the characteristic features of the proposed scheme for state generation are marked by simplicity (just linear optics and post-selection of the coincidence type), limited initial resources (just three independent particles in the input), and universal efficiency (the same for any desired state). This makes the proposal an interesting technique for integrated quantum technologies motivating further research towards extension to arbitrary number of qubits and improvement of the efficiency of the scheme.

\vspace{0.1cm}

\textit{Acknowledgments.}---We thank Yong-Su~Kim and Marek~\.Zukowski for discussions and helpful comments. MM acknowledges partial support by the Foundation for Polish Science (IRAP project, ICTQT, contract no. MAB/2018/5, co-financed by EU within Smart Growth Operational Programme).

\bibliography{CombQuant}

\end{document}